\documentclass[11pt]{article}
\usepackage[paper=a4paper,dvips,top=2.5cm,left=2.5cm,right=2.5cm,bottom=2.6cm,footskip=2.2cm,voffset=0.0cm, marginparwidth=18mm]{geometry}
\usepackage{amsfonts,amsmath,amsthm,amssymb}
\numberwithin{equation}{section}
\usepackage[svgnames]{xcolor}
\usepackage{fix-cm}
\usepackage{sectsty}
\usepackage{fancyhdr}
\usepackage{lastpage}
\usepackage{graphicx}
\usepackage{url}
\usepackage{enumerate}
\usepackage{paralist}
\usepackage{multicol}
\usepackage{color}
\usepackage{float}
\usepackage{epsfig}
\usepackage{subfigure}
\usepackage{hyperref}
\hypersetup{colorlinks=true,linkcolor=beamer@PRD, citecolor=beamer@PRD}
\usepackage{authblk}
\usepackage{cite}
\usepackage{todonotes}
\usepackage{ulem} % Enables sout command
\renewcommand\eqref[1]{\textcolor{beamer@PRD}{(}\ref{#1}\textcolor{beamer@PRD}{)}}
\definecolor{beamer@PRD}{RGB}{46,48,146}
\begin{document}
%%%%%%%%%%%%%%%%%%%%%%%%%%%%%%%%%%%%%%%%%%%%%%%%%%%%%%%%%%%%%%%%%%%%%%%%%%%%%%
%  Title
%%%%%%%%%%%%%%%%%%%%%%%%%%%%%%%%%%%%%%%%%%%%%%%%%%%%%%%%%%%%%%%%%%%%%%%%%%%%%%

\title{ Consequences of Gödel Theorems  on Third Quantized Theories Like String Field Theory and Group Field Theory }
\author{\small{Mir Faizal$^{1,2,3}$, Arshid Shabir$^1$, Aatif Kaisar Khan$^{1,5}$}
\\
\textit{\small $^{1}$Canadian Quantum Research Center, 204-3002, 32 Ave Vernon, BC V1T 2L7, Canada}\\
\textit{\small $^{2}$Irving K. Barber School of Arts and Sciences, University of British Columbia Okanagan, Kelowna, BC V1V 1V7, Canada}\\

\textit{\small $^{3}$Department of Mathematical Sciences, Durham University,
Upper Mountjoy,  Stockton Road, Durham DH1 3LE, UK}\\ 
\textit{\small $^{5}$Facultat de Física, Universitat de Barcelona, E-08028 Barcelona, Spain}
}
\date{}
\maketitle
%%%%%%%%%%%%%%%%%%%%%%%%%%%%%%%%%%%%%%%%%%%%%%%%%%%%%%%%%%%%%%%%%%%%%%%%%%%%%%
\begin{abstract}
The observation that spacetime and quantum fields on it have to be dynamically produced in any theory of quantum gravity implies that quantum gravity should be defined on the configuration space of fields rather than spacetime. Such a theory is described on the configuration space of fields rather than spacetime, which is a third quantized theory. So, both string theory and group field theory are third-quantized theories. Thus, using axioms of string field theory, we motivate similar axioms for group field theory.
Then, using the structure of these axioms for string field theory and group field theory, we identify general features of axioms for any such third quantized theory of quantum gravity. Thus, we show that such third-quantized theories of quantum gravity can be formulated as formal axiomatic systems.  We then analyze the consequences of Gödel theorems on such third quantized theories.  We thus address problems of consistency and completeness of any third quantized theories of quantum gravity.
\end{abstract}

\tableofcontents
\newpage

\section{Introduction}

 The structure of spacetime can be obtained from general relativity. An intriguing aspect of general relativity is that it predicts its own breakdown due to the occurrence of singularities. At singularities, the spacetime description of reality fails, as the curvature of spacetime becomes infinite, and the laws of physics as described by general relativity cease to be valid. The Penrose-Hawking singularity theorems reveal that these singularities are inherent to the very structure of general relativity \cite{Penrose1965, Hawking1970}. Thus, the breakdown of the spacetime description of physics is intrinsic to the very nature of spacetime as described by general relativity.
 
Quantum gravitational effects are expected to modify this classical description of spacetime, incorporating a natural geometric cutoff that prevents the formation of singularities \cite{Garay1995, Melchor:2023rqd, Blanchette:2021vid, Rastgoo:2022mks}.
 In string theory, for instance, T-duality introduces a minimal length scale, below which the conventional notions of spacetime cease to exist \cite{Polchinski1998, Lust1989}. This minimal length effectively prevents the occurrence of singularities by ensuring that physical quantities remain finite \cite{Jusufi:2023pzt, Brandenberger:2018xwl, Bossard:2002ta}. 
 Such a geometric cutoff also occurs in Loop Quantum Gravity (LQG) due to the discrete nature of the theory 
\cite{Ashtekar2005, Rovelli2004}.
In loop quantum cosmology (LQC), singularities are avoided due to a discrete structure of spacetime, which introduces a geometric cutoff \cite{Ashtekar2006, Bojowald2005}. The application of LQC to early universe cosmology has demonstrated that quantum geometric effects can resolve the Big Bang singularity, replacing it with a quantum bounce \cite{Bojowald2001, Yang:2022aec}. 
These findings suggest that the absence of singularities is a universal feature of any consistent theory of quantum gravity. 

This absence of singularities can also be obtained using the Bekenstein-Hawking entropy \cite{Bekenstein1973, Hawking1975}. The modification to the Bekenstein-Hawking entropy by a geometric cutoff would naturally prevent the formation of singularities, ensuring that the physical description remains finite and well-defined. 
The Jacobson formalism further strengthens this connection by directly linking the Bekenstein-Hawking entropy to spacetime geometry \cite{Jacobson1995}. Modifications to this entropy, as predicted by quantum gravity theories, consequently alter the underlying geometry of spacetime. Explicit demonstrations show that the bound on Bekenstein-Hawking entropy due to a minimal length in quantum gravity prevents the formation of spacetime singularities \cite{Awad:2014bta, Salah:2016kre}. These results suggest that singularities arise in general relativity when it is applied to regimes where the spacetime description becomes invalid. Importantly, this geometric bound is derived from a bound on quantum information, suggesting that spacetime geometry may emerge from quantum informational principles \cite{Cai2010, Horowitz1996}.
This implies that in quantum gravity, spacetime is not a fundamental entity but an emergent phenomenon arising from a more fundamental quantum theory \cite{mi12, mi14}. Various approaches to quantum gravity, such as string theory and LQG, indicate the necessity of a third quantized theory to explain such dynamical formation of spacetime and geometric structures \cite{Gielen2016, Oriti2016, Oriti2017}.

In a first quantized theory, the quantum mechanics of individual particles are studied, while in a second quantized theory, the quantum mechanics of fields are considered, allowing for the dynamic creation and annihilation of particles. Second quantization thus naturally explains multi-particle systems, where the wave function is defined on the configuration space of fields rather than spacetime. This leads to the concept of third quantization, where quantum theory is constructed on this abstract configuration space of fields. Consequently, both quantum fields and the underlying geometry on which these fields are defined are dynamically created and annihilated in a third quantized theory.
A third quantized theory is not constructed within spacetime; rather, spacetime and quantum fields emerge from it. Third quantization is thus a multi-geometry theory, analogous to how second quantization is a multi-particle theory. The third quantization of the Wheeler-DeWitt approach has yielded various interesting results \cite{ Kiefer2012, Kuchar1992}. For instance, the application of third quantization to quantum cosmology has provided insights into the creation and annihilation of universes in a multiverse  \cite{Hartle1983, Vilenkin1983}. Thus, using a third quantization, not only the emergence of a single universe but an entire multiverse can be explained.  

The third quantization of LQG has been studied using  Group Field Theory (GFT), which provides a field-theoretic formulation of quantum geometry \cite{Oriti2017}. GFT describes quantum states of geometry using group-theoretic variables, allowing for a combinatorial and algebraic approach to quantum gravity. This framework has been instrumental in understanding the dynamics of quantum spacetime and the transition from quantum to classical geometry \cite{Freidel2005, Oriti2011}. Moreover, GFT has been connected to spin foam models, which serve as a covariant formulation of LQG, further bridging the gap between canonical and path integral approaches \cite{Perez2003}.
Similarly, in string theory, the String Field Theory (SFT) is also a third quantized theory. It may be noted that historically SFT  is sometimes called a second quantized theory. This is because string theory can be viewed as either a first quantized theory of strings or a second quantized conformal field theory. 
Thus, SFT  can be seen as a third quantized theory defined on the configuration space of the conformal field theory or equivalently as a second quantized theory of strings. So,
despite being termed a second quantized theory, SFT  operates on the configuration space of fields, fitting the criteria of a third quantized theory \cite{Siegel1999, Hata1986}. SFT  provides a consistent framework for describing the interactions of strings, incorporating both perturbative and non-perturbative effects. This approach has led to significant insights into the non-perturbative structure of string theory, including the study of D-branes, tachyon condensation, and string dualities \cite{Sen1999, Zwiebach1993}.

The universe/multiverse with quantum fields in it emerges from such a third quantized theory of quantum gravity. 
As such third quantized theories are not defined in spacetime, but rather spacetime emerges from them. The third quantization exists as an axiomatic structure producing spacetime and quantum field from it. Now consistency and completeness are the bare minimum requirement for any such sensible theory, describing physics at a fundamental level. The theory should not produce contradicting results, as then it would not be a sensible theory describing reality. Furthermore, as this theory is a fundamental theory, it should be complete, and all physical phenomena that can occur in nature should be derivable from it. However,  the   Gödel first and second theorems  \cite{12, 12a} have direct implications for the construction of such a theory. We will analyze such formal aspects of third quantized quantum gravity in this paper.  

\section{Third Quantized Theories}
Even though there are several approaches to quantum gravity, the requirement to dynamically create and annihilate geometries seems to naturally lead to some sort of third-quantized field theory. It may be noted that the original works on third-quantized field theories were done in the context of the Wheeler-DeWitt equation \cite{Giddings}. As canonical quantum gravity, based on the   Wheeler-DeWitt equation, evolved to 
LQG \cite{Thiemann:2002nj}, the third quantized Wheeler-DeWitt equation evolved to GFT. The GFT is a higher-dimensional extension of matrix models and so it provides a developed formalism for third-quatized LQG. In GFT, the fundamental entities are fields defined on a group manifold, corresponding to the quantized geometric degrees of freedom in LQG. The GFT framework encodes the dynamics of spin networks, the basic quantum states in LQG. A typical GFT action is given by:
\begin{equation}
S_{\text{GFT}} =\frac{1}{2} \int \prod_{i=1}^D \mathrm{d}g_i \, \Psi(g_1, \ldots, g_D) \mathcal{K}_G  \Psi(g_1, \ldots, g_D) + \lambda \int \prod_{i=1}^D \mathrm{d}g_i \, \mathcal{V}(\Psi(g_1, \ldots, g_D))
\end{equation}
where \( g_i \) are elements of a Lie group \(\mathcal{G}\), \( \Psi(g_1, \ldots, g_D) \) is a field over \( D \) copies of the group \(\mathcal{G}\), \( \mathcal{K}_G \) is a kinetic term, and \( \mathcal{V} \) represents the interaction term \cite{Freidel2005, Oriti2006, Oriti2012}.
 The interaction terms in the action often correspond to combinatorial structures, such as simplices or graphs, encoding the connectivity and topology of the fields. This reflects the discrete nature of spacetime in these theories \cite{Baratin2012}.
In third quantized LQG, the field \( \Psi(g_1, \ldots, g_D) \) can be interpreted as creating and annihilating quantum geometries, with the kinetic and interaction terms encoding the dynamics of these geometries. The fields in GFT are analogous to wave functions over the configuration space of spin networks, encapsulating the dynamics of LQG in a field-theoretic formalism \cite{Oriti2011}.

String Field Theory (SFT) provides a third quantized description of string theory, where the basic objects are string fields that can create and annihilate strings. Even though it has been historically viewed as a second quantized theory of strings, it can also be equivalently viewed as field theory defined on conformal fields and hence can be seen as a third quantized theory. The SFT action encapsulates the dynamics of these string fields and includes terms corresponding to the free propagation of strings and their interactions. In string theory, a covariant open bosonic string field theory stands as a significant milestone, offering a powerful framework for understanding the dynamics of open strings. This is constructed using  an action that resembles the action of the Chern-Simons theory,
\begin{equation}\label{action}
S= \frac{1}{2}\int \Psi \star Q\Psi + \frac{\lambda}{3}\int\Psi \star\Psi \star\Psi,
\end{equation}
The action embodies the stringy dynamics, and the coupling constant \(\lambda\) controls the strength of interactions. Through this action, strings are endowed with a rich algebraic structure \( \mathcal{A} \), governed by a non-commutative star product \( \star: \mathcal{A} \otimes \mathcal{A} \rightarrow \mathcal{A} \) that encapsulates the gluing of incoming strings into composite entities. Furthermore, the action incorporates a BRST operator \( Q: \mathcal{A}\rightarrow \mathcal{A} \) reflecting the underlying symmetries of the string worldsheet \cite{Siegel1988, Zwiebach1993, Kostelecky1989}.

The string field \( \Psi(X) \) encompasses all possible string configurations, and the action \( S_{\text{SFT}} \) describes how these configurations evolve and interact. The BRST operator \( Q \) ensures that string field theory is gauge-invariant, and the interaction term \(\Psi \star\Psi \star\Psi \) represents the merging and splitting of strings. It may be noted that strings have branes, and even though attempts to construct brane field theories have been made, it has not been possible to construct a fully developed brane field theory. So, brane can be seen as derived objects in string theory, rather than fundamental objects from which an independent theory can be constructed \cite{W}. 

So, the concept of third quantization extends the idea of second quantization to a field-theoretic setting where the fields themselves represent quantum states of a system, such as spin networks in LQG or strings in SFT. The general structure of third quantized field theories can be described using the analogy with SFT and GFT. 
The fields are defined over a configuration space that represents the quantum states of the underlying theory (e.g., spin networks for LQG, string configurations for SFT).
The kinetic term in the action describes the free propagation of these fields. It typically takes the form \( \int \Psi \mathcal{K} \Psi \), where \( \mathcal{K} \) is an appropriate operator, and depends on the theory. In GFT,   \( \mathcal{K} = \mathcal{K}_G \) and in SFT \( :\mathcal{K} = Q \). 
The interaction term describes the interactions between the fundamental quantum states. It usually involves higher-order products of the fields, such as \( \int \Psi^n \), where the product of fields depends on the details of the theory. 
Thus, action for any third quantized field theory can be generically written as:
\begin{equation}
S =\int \mathcal{D}\Psi \left( \frac{1}{2} \Psi \mathcal{K} \Psi + \mathcal{V}(\Psi) \right)
\end{equation}
where \( \mathcal{K} \) is the kinetic operator, \( \mathcal{V}(\Psi) = \Psi^n \) represents the interaction term, and \( \lambda \) is a coupling constant. Here, this product of third quantized fields, and the Kinetic term depends on the nature of the theory. The fields can form superpositions of different configurations, allowing for the exploration of a wide range of possible states.
 The dynamics can be formulated using a path integral over the fields, integrating over all possible configurations.
Now any such theory can be third quantized using a path integral over all possible field configurations:
\begin{equation}
\mathcal{Z} = \int \mathcal{D}\Psi \, e^{i S(\Psi)}
\end{equation}
which defines the partition function and encapsulates the quantum dynamics of the theory. 

Here, it may be noted that the third quantized field is not defined over spacetime but field configurations, the details of which depend on the exact nature of the third quantized theory. So, the spacetime representing the universe/multiverse, and the quantum fields in it will emerge as emergent phenomena from such a third quantized theory \cite{mi12, mi14}. Hence, such a third quantized theory does not exist in spacetime, but rather it exists in a Platonic realm. 
Here, the term Platonic realm is borrowed from 
philosophical theory, and it posits that the material world is not the true reality, but rather a shadow of the true reality, which consists of abstract, non-material forms or ideas \cite{P1, P2}. Here, in quantum gravity, this world of abstract, non-material forms is represented by the axiomatic structure of third-quantized theories. The Platonic nature of modern physics has already been discussed \cite{P}, what is interesting about third quantized quantum gravity is that even spacetime is an emergent phenomenon in it. Now to analyze the consequences of this further it is important to first investigate this concept of third quantization as a formal axiomatic system.

\section{String Field Theory and Group Field Theory}  

In this section, we will review the axioms of  SFT \cite{zwiebach1992, taylor2004, Witten1986}, and use them to motivate similar axioms for GFT. This is important to
 understand how third quantized theories can be viewed as a formal system. 
Now it is known that Witten's string field theory is a formulation of string theory that describes the dynamics of strings using a field theory approach and can be analyzed as a formal axiomatic system with the following axioms \cite{zwiebach1992, taylor2004, Witten1986}: 

1. \textbf{String Field}: The string field \( \Psi \) is a ``functional" of the string's configuration $X$. This is encoded in the position and momentum of the string, or equivalently in the conformal field theory language by vertex operators. It has all the modes of a string which account for its degrees of freedom.

2. \textbf{Inner Product}:   Integration over the world sheet of the string gives the inner product \( \langle \cdot, \cdot \rangle \), this ensures that action is a scalar quantity and one can obtain it by integrating over all possible configurations of strings $\int DX$. 

3. \textbf{BRST Invariance}: The action is invariant under BRST transformations which guarantee gauge symmetry in a theory. The BRST operator \( Q \) encodes the constraints and symmetries of the theory. Physical states are identified as cohomology classes of \( Q \), meaning they satisfy \( Q \Psi = 0 \) and are not exact, i.e., \( \Psi \neq Q \chi \) for any \( \chi \)~\cite{brst1976}.

4. \textbf{Star Product}: The star product \( \star \) is a non-commutative product on the space of string fields $  \Psi_1 \star \Psi_2\neq \Psi_2 \star \Psi_1$. It represents the interaction of strings, and so captures the joining and splitting of strings.

5. \textbf{Gauge Invariance}: The action is invariant under a set of gauge transformations of the form:
   \begin{equation}
   \delta \Psi = Q \Lambda + \Psi \star \Lambda - \Lambda \star \Psi,
   \end{equation}
   where \( \Lambda \) is a gauge parameter.

6. \textbf{Associativity}: In string field theory, the star product $\star$, which defines the interaction between string fields, must satisfy the associativity. This property ensures that the product of three string fields is independent of the order of operations. Mathematically, this can be expressed as:
\begin{equation}
(\Psi_1 \star \Psi_2) \star \Psi_3 = \Psi_1 \star (\Psi_2 \star \Phi_3)
\end{equation}
where the fields $\Psi_1$, $\Psi_2$, and $\Phi_3$ represent strings fields. 

7. \textbf{Action Principle}: The theory is governed by an action \( S \), which is a ``functional" of a string field \( \Psi \). The action for bosonic string field theory is given by:
   \begin{equation}
   S(\Psi) = \frac{1}{2} \langle \Psi, Q \Psi \rangle + \lambda \mathcal{V}(\Psi) 
   \end{equation}
   where  $\mathcal{V}(\Psi)  = \langle \Psi, \Psi \star \Psi \rangle/3$, 
   \( \langle \cdot, \cdot \rangle \) denotes an inner product on the space of string fields and \( Q \) is the BRST operator, \(\lambda\) is the string coupling constant, and \( \star \) is the star product~\cite{Witten1986}.

Here, we note that the star product $\star$ is a non-commutative product that encodes the interaction rules for the string fields. It reflects the physical process of joining and splitting strings. The associativity property ensures that the product of three string fields is well-defined and consistent, regardless of how the fields are grouped. This would be true even in other third-quantized theories, like GFT. As GFT is also a third quantized theory \cite{Freidel2005, Oriti2006}, we can use the axioms of Witten's SFT, to motivate the axioms for GFT. Thus, using the work done on SFT, and the properties of GFT, it is possible to propose the following axioms for GFT:

1. \textbf{Field on Group Manifold}: The fundamental variables are fields \( \Psi(g_1, g_2, \ldots, g_n) \) defined on a group manifold, where \( g_i \) are elements of a Lie group \({G}\) \cite{Oriti2007}.

2. 
\textbf{Inner Product}: 
Given two group fields \(\Psi_1\) and \(\Psi_2\), the inner product \(\langle \Psi_1, \Psi_2 \rangle\) is generally defined as:
\begin{equation}
\langle \Psi_1, \Psi_2 \rangle = \int \prod_{i=1}^d dg_i \, {\Psi_1(g_1, g_2, \ldots, g_d)} \Psi_2(g_1, g_2, \ldots, g_d)
\end{equation}
where \(d\) is the number of group elements associated with the field and 
\(dg_i\) is the Haar measure on the group \(G\), which ensures that the integral is invariant under group transformations.

3. \textbf{Gauge Invariance}: The theory is invariant under local gauge transformations of the fields, typically under the action of the group \({G}\) ({for \( h \in G \)} \cite{Freidel2005}):
   \begin{equation}
   \Psi(g_1, g_2, \ldots, g_n) \rightarrow \Psi(h g_1, h g_2, \ldots, h g_n)
   \end{equation}

4. \textbf{ Star Product}: In  GFT, the interaction can also be defined using a \(\star\)-product, which is a noncommutative product that combines group fields. For three group fields \(\Psi_1\), \(\Psi_2\), and \(\Psi_3\) defined on the group \(G\), the star product is defined as \cite{Freidel2005}:
\begin{equation}
(\Psi_1 \star \Psi_2 \star \Psi_3)(g_1, g_2, g_3, g_4) = \int dh \, \Psi_1(g_1, g_2, h) \Psi_2(h^{-1}, g_3, g_4) \Psi_3(g_4^{-1}, h)
\end{equation}
 The fields \(\Psi_1\), \(\Psi_2\), and \(\Psi_3\) are functions defined on the group manifold \(G\). Each field depends on three group elements, for instance, \(\Psi(g_1, g_2, g_3)\). The integral \(\int_G dh\) is taken over the group \(G\) concerning the Haar measure \(dh\), which ensures invariance under group transformations.  
 
5. \textbf{Symmetry and Invariance}: The action should respect the symmetries of the underlying group manifold, such as rotational and Lorentz invariance for relevant physical applications.
 
6. \textbf{Associativity}: The interaction terms should respect an associative product structure, analogous to the star product in SFT, to ensure the consistency of the interactions, 
\begin{equation}
(\Psi_1 \star \Psi_2) \star \Psi_3 = \Psi_1 \star (\Psi_2 \star \Phi_3)
\end{equation}
where the fields $\Psi_1$, $\Psi_2$, and $\Phi_3$ represent fields in GFT.

7. \textbf{Action Principle}: The dynamics of the fields are governed by an action \( S \), which is a ``functional" of the group fields. The action typically includes kinetic and interaction terms:
\begin{equation}
S_{\text{GFT}} = \frac{1}{2}\langle \Psi, \mathcal{K}_G \Psi\rangle  + \lambda \mathcal{V}(\Psi)
\end{equation}
   where \( \mathcal{K}_G \) is the kinetic operator and \( \mathcal{V} (\Psi)\) represents interaction terms \cite{Oriti2011}. Here, $\mathcal{V}(\Psi)$ also includes an integration involving the Haar measure and is constructed using the star product for GFT. 

Here again, the associativity is crucial for the internal consistency of GFT, as it ensures that interactions are unambiguously defined. Thus, motivated by axioms of SFT, it was possible to suggest similar formal axioms for GFT.

\section{Gödel's Theorems Applied to Third Quantized Theories}
As both SFT and GFT are third-quantized theories, we can use them to understand the general structure of any third-quantized theory. It may be noted that even if LQG or string theory is not the theory of quantum gravity,   any theory should produce spacetime and quantum fields on spacetime dynamically. Thus, it would be a theory defined on the configuration space of fields rather than spacetime and hence a third quantized theory. Now for any such theory, we can identify its general features using SFT and GFT. Thus, 
using  the structure of SFT and GFT, we infer some general features of such an axiomatic system as follows:  

1. \textbf{Fields on Configuration Space}
The fundamental objects are fields defined on a configuration space  \(\chi\) that represents the degrees of freedom of the theory. For example, in SFT, these are string fields $\Psi[X]$ where $X$ represents the string configuration \cite{Witten1986}, and in GFT, these are group fields $\Psi(g_1, g_2, \ldots, g_n)$ \cite{Oriti2011}. 

2. \textbf{Inner Product and Hilbert Space}
There is a well-defined inner product on the space of fields $
\langle \Psi, \Psi \rangle
$,  ensuring that the action is a scalar quantity. 
This inner product induces a Hilbert space structure on the space of states and can be defined by integrating over the configuration space on which the third quantized field is defined. 

3.  \textbf{Gauge Invariance}: The field \(\Psi(\chi)\) is subject to gauge transformations that ensure the invariance of the action. These transformations depend on the specific symmetries of the configuration space \(\chi\).

  4.   \textbf{Interaction Terms}: The interaction terms \(\mathcal{V}(\Psi)\) describe the interactions between the fields. These terms are constructed to respect the symmetries of the theory and involve higher-order products of the fields.  To construct such interaction terms, we can define an associative \(\star\) product, which is generally defined in the field space. For two fields \(\Psi_1\) and \(\Psi_2\), the \(\star\) product \(\Psi_1 \star \Psi_2\) is given by:
\begin{equation}
(\Psi \star \Psi)(\chi) = \int d\chi_1\, d\chi_2 \, K(\chi, \chi_1, \chi_2) \Psi(\chi_1) \Psi(\chi_2)
\end{equation}
where \(\chi\), \(\chi_1\), and \(\chi_2\) denote third quantized fields, and 
\(K\) is a kernel that encodes the interaction rules.   
  
  5.   \textbf{Kinetic Term}: The kinetic term \(\mathcal{K}\) governs the free propagation of the fields and typically involves appropriate operators operating in third-quantized fields. For SFT, this would be the BRST operator $\mathcal{K} = Q$, and for GFT this would be $\mathcal{K} = \mathcal{K}_G$.

6. \textbf{Associativity}: The interaction terms should respect an associative product structure to ensure the consistency of the theory. This is analogous to the star product in SFT and the product structure in GFT.

7. \textbf{Symmetry and Invariance}
The action and the theory respect the symmetries of the underlying configuration space, such as the Lorentz invariance in SFT \cite{Witten1986} and the symmetries of the group manifold in GFT \cite{Oriti2011}.
 
 8. \textbf{Action Principle}
The dynamics of the fields are governed by an action $S$, which is a ``functional" of these fields. The action typically includes kinetic and interaction terms:
\begin{equation}
S(\Psi) = \frac{1}{2} \langle \Psi, \mathcal{K} \Psi \rangle + \lambda \mathcal{V}(\Psi)
\end{equation}
where $\Psi$ represents the field, $\mathcal{K}$ is the kinetic operator, $\mathcal{V} (\Psi) $ represents interaction terms, and $\langle \cdot, \cdot \rangle$ denotes an appropriate inner product. Here, $\mathcal{V} (\Psi) $ (like the inner product) would also include an integration over the configuration space of the third quantized theory. The details of this world depend on the specifics of the theory. The strength of the interaction is controlled by the coupling constant $\lambda$.  
 
Now we observe that third quantized theories are basically consistent formal systems \( \mathcal{F} \), which are not present in spacetime, but rather spacetime emerges from them as their consequences. Thus, they exist in some Platonic realm, and in that Platonic realm apart from that formal system \( \mathcal{F} \), there also exists a computation algorithm \( \mathcal{C} \) to derive the corollaries of that system \( \mathcal{F} \). The spacetime along with quantum fields on it emerges as the corollaries of that system \( \mathcal{F} \).  For SFT, this system is represented by the axioms of SFT and axioms of quantum mechanics to third quantize it, and similarly for GFT, this structure is represented by axioms of GFT and axioms of quantum mechanics. It may be noted that quantum mechanics can be viewed as an axiomatic structure \cite{Sakurai, Ballentine,Nielsen}. Using these axioms of quantum mechanics, we could possibly define an operator algebra for any third-quantized theory of quantum gravity. 
Thus, the theory should include an algebra of operators that create and annihilate configurations, reflecting the third quantization process. These operators satisfy commutation or anti-commutation relations depending on the nature of the third field. The spacetime and quantum fields on it would emerge as emergent phenomena using these operators. 
 It is possible that the final theory could modify the operator algebra too due to quantum gravitational effects. This has already been proposed in objective collapse models, where gravitational effects modify the quantum mechanics and cause a scale-dependent collapse of the wave function \cite{Bassi2003, Penrose1994}. In fact, such modification of quantum mechanics has been applied to the second quantized Wheeler-DeWitt equation, and it resolves certain problems associated with the usual Wheeler-DeWitt equation  \cite{j1, j2}. It would thus be possible to generalize this work to third-quantized quantum gravity. Thus, along with the axioms of quantum mechanics (or its suitable modifications \cite{Bassi2003, Penrose1994}), third-quantized quantum gravity can be viewed as an axiomatic system   \( \mathcal{F} \). The spacetime will be an emergent phenomenon from it, and so it cannot be possibly defined in spacetime. Thus, this system will exist in the Platonic realm and not spacetime. 

Gödel's incompleteness theorems \cite{12, 12a} will now apply to third quantized theories, as it is represented by a formal system \( \mathcal{F} \) that exists in a Platonic realm. Two of the most important findings in mathematical logic are Gödel's incompleteness theorems, which show that formal axiomatic systems that can represent elementary arithmetic have intrinsic limits.
Now, as any third quantized theory can be viewed as such a formal axiomatic system \( \mathcal{F} \), Gödel's incompleteness theorems will be applicable to them. Thus, if we start by constructing a consequence \( \mathcal{G} \) within \( \mathcal{F} \) such that \( \mathcal{G} \) asserts its own unprovability, we have $
\mathcal{G} \equiv $``This statement is not provable in $ \mathcal{F} $.''
In the formal system \( \mathcal{F} \), there exists a sentence \( \mathcal{G} \) such that
$
\mathcal{F} \nvdash \mathcal{G} \quad \text{and} \quad \mathcal{F} \nvdash \neg \mathcal{G}.
$
This means \( \mathcal{G} \) is true but unprovable within \( \mathcal{F} \). Gödel's Second Incompleteness Theorem further states that, let \(\text{Con}(\mathcal{F})\) be the statement within \( \mathcal{F} \) that asserts the consistency of \( \mathcal{F} \):
$
\text{Con}(\mathcal{F}) \equiv \text{``There is no statement } \varphi \text{ such that both } \varphi \text{ and } \neg \varphi \text{ are provable in } \mathcal{F} \text{.''}
$
The formal system \( \mathcal{F} \) cannot prove its own consistency, which is
$
\mathcal{F} \nvdash \text{Con}(\mathcal{F}),$ 
If \( \mathcal{F} \) is consistent, then \(\text{Con}(\mathcal{F})\) is true, but \(\text{Con}(\mathcal{F})\) is not provable within \( \mathcal{F} \).

Here, third quantized fields, such as string fields or fields in GFT, \(\Psi\) and related operators/functions play the role of formal axioms and rules of inference. The statements within the formal system \( \mathcal{F} \) correspond to possible configurations and interactions of the third quantized field \(\Psi\) and its associated operations. Arithmetic statements are encoded within this framework, representing numbers and arithmetic operations using the structures in the system.

Let \( \mathcal{F} \) be the formal system derived from any third quantized theory, such as SFT or GFT. The Gödel Sentence \(\mathcal{G}\) is defined as
$\mathcal{G} \equiv$ {``This configuration cannot be derived from the axioms of } $ \mathcal{F}$.
This means that \( \mathcal{F} \nvdash \mathcal{G} \) and \( \mathcal{F} \nvdash \neg \mathcal{G} \). Moreover, $\text{Con}(\mathcal{F})$ is defined as a consistency statement
$\text{Con}(\mathcal{F})\equiv $. There exists no configuration $ \varphi $ such that both $\varphi$ and $ {\neg \varphi} $ are derivable from the axioms of $\mathcal{F}$.
Therefore we have \( \mathcal{F} \nvdash \text{Con}(\mathcal{F}) \).
These expressions cover the use of Gödel’s incompleteness theorems within a formal system representing any third quantized theory. Thus, it reveals limits in proving some statements and self-consistency of this system.
The third quantized theories which are formal systems  \( \mathcal{F} \) existing in the Platonic realm will thus be subject to Gödel’s incompleteness theorem.  

\section{A Consistent and Complete Third Quantized Theory}
Now the problem with this application of the Gödel’s incompleteness theorem in the Platonic realm is that it is applied to the actual axiomatic structure \( \mathcal{F} \) describing reality rather than 
human understanding of \( \mathcal{F} \) (which we will denote by $F$). It is possible to have inconsistencies in $F$, but is by definition impossible to have inconsistencies in \( \mathcal{F} \) and as \( \mathcal{F} \).  Furthermore, all physical phenomena in the universe/multiverse are obtained as corollaries of \( \mathcal{F} \) using some computational algorithm \( \mathcal{C} \), which also exists in the Platonic realm. However, there are things which are true due to the very structure of \( \mathcal{F} \), but cannot be obtained using any 
 computational algorithm \( \mathcal{C} \). Thus, something more is required in the Platonic realm other than   \( \mathcal{F} \) and \( \mathcal{C} \) to resolve this problem. 
  Now it may be noted that the Lucas-Penrose argument has addressed the Gödelian limitations in $F$. 
The Lucas-Penrose argument contends that mechanical systems (such as computers or formal systems) are unable to accurately represent human minds, using Gödel's incompleteness theorems \cite{Lucas1961, Penrose1989, Penrose1994}. It explains that to explain how the human mind can overcome the Gödelian limitations, and see the validity of the Gödelian statement.  Thus, for humans, if they identify $F$, and also identify a computational algorithm $C$ (which will correspond to human understanding of \( \mathcal{C} \)) to derive corollaries of $F$, they will also have a non-computational non-algorithmic understanding denoted by $N$, which will overcome the Gödelian limitations. Using $N$ they can see the validity of Gödelian statements.  

Now we generalize this original  Lucas-Penrose argument to the Platonic realm, and so corresponding to $N$, we define a non-computational non-algorithmic understating  in the Platonic realm as   \( \mathcal{N} \). 
Now just as for the formal systems $F$,    $C$ cannot validate Gödelian consequences, and this can only be done by $N$ in human brain.   For the formal systems $\mathcal{F}$ in the Platonic realm,    $\mathcal{C}$ cannot validate its Gödelian consequences, and this can only be done by $\mathcal{N}$.  Here, both $\mathcal{C}$  and $\mathcal{N}$ also exist in the Platonic realm.  
So, just as non-algorithmic understanding in the human brain is needed to obtain a complete consistent understanding of reality by humans,  non-algorithmic understanding in the Platonic realm is needed to actualize a complete consistent description of reality. This is the only way to avoid inconsistencies and incompleteness in the universe/multiverse. 
If we were to restrict our framework to \( \mathcal{F} \) and \( \mathcal{C} \), it would be impossible to achieve a complete consistent description of reality. Certain truths derived from \( \mathcal{F} \) would remain unprovable within \( \mathcal{C} \). The resolution to this issue lies in generalizing the original Lucas-Penrose argument by incorporating \( \mathcal{N} \) into the Platonic realm. This inclusion of this non-algorithmic understanding in the Platonic realm is the only way to overcome the Gödelian limitations inherent in \( \mathcal{C} \).
It may be noted like  \( \mathcal{F} \) and  \( \mathcal{C} \),  this \( \mathcal{N} \) also operates in the Platonic realm and should not be confused with the human understanding of  \( \mathcal{N} \) (denoted by $N$).  Thus, we will now apply the argument in an abstract setting to the actual theory in the Platonic realm, where any computation performed using the formal system \(\mathcal{F}\) is denoted as \(\mathcal{C}\), and any conclusion derived in a non-algorithmic non-computational way is denoted as \(\mathcal{N}\). Gödel's incompleteness theorems apply if they remain restricted to \(\mathcal{C}\). However, in the Platonic realm, we will see how the generalization of the Lucas-Penrose argument resolves this difficulty.   

To mathematically express how this argument might overcome the limitations in a formal system based on any third quantized theory of quantum gravity, we need to illustrate how \(\mathcal{N}\) identifies truths that the formal system  \( \mathcal{F} \),  cannot prove in the Platonic realm using  \( \mathcal{C} \). Let \(\mathcal{G}\) be a Gödel sentence in a formal third quantized system, \(\mathcal{F}\):
$
\mathcal{G} \equiv $``This statement is not provable in $ \mathcal{F} $.''
According to Gödel's first incompleteness theorem,
$
\mathcal{F} \nvdash \mathcal{G} \quad \text{and} \quad \mathcal{F} \nvdash \neg \mathcal{G}.
$ Lucas-Penrose argument says that if \(\mathcal{N}\) is used, it can be seen that \(\mathcal{G}\) is true even if \(\mathcal{F}\) cannot prove \(\mathcal{G}\). Formally, one can say by using \(\mathcal{N}\) that the truth of \(\mathcal{G}\) is beyond the formal system \(\mathcal{F}\). This understanding allows us to enlarge the formal system \(\mathcal{F}\) into another system \(\mathcal{F}^{\prime}\) whose addition as an axiom in it is \(\mathcal{G}\):
$
\mathcal{F}^{\prime} = \mathcal{F} + \{ \mathcal{G} \}
$.
To verify this, we suppose
$
\text{if } \mathcal{F} \text{ is consistent, then } \mathcal{F}^{\prime} \text{ is consistent}
$. 
To summarize mathematically, we express the overcoming of limitations as follows:
$
\mathcal{G} \equiv $``This statement is not provable in  $\mathcal{F}$.''
So,
$
\mathcal{F} \nvdash \mathcal{G} \quad \text{and} \quad \mathcal{F} \nvdash \neg \mathcal{G}
$. Essentially, such a generalization on Lucas-Penrose's argument suggests that the limitations imposed by Gödel's theorems in a formal system based on third quantized theory can be transcended through a non-algorithmic \(\mathcal{N}\).

The original Lucas-Penrose argument based on $F, C, N$ has drawn its own criticisms. 
The argument assumes that a single formal system can encapsulate human reasoning. But informal thinking that is not captured by a set formal system or a succession of changing formal systems may be involved in human reasoning \cite{Searle1992}.  Furthermore, the argument is predicated on the consistency of the formal system that models human reasoning. The argument falls apart if human reasoning is inconsistent \cite{Chalmers1996}. 

These criticisms of the original Lucas-Penrose argument aim at the human understanding of \( \mathcal{F} \), \( \mathcal{C} \), \( \mathcal{N} \) i.e. $F, C, N$,  rather than at the actual  \( \mathcal{F} \), \( \mathcal{C} \), \( \mathcal{N} \) in the Platonic realm.
For  \( \mathcal{F} \),  in the Platonic realm, as the universe/multiverse exists as its consequence through \( \mathcal{C} \),  then we need an \( \mathcal{N} \) in the Platonic realm too, to avoid problems due to Gödel's theorems.  It is consistent to acknowledge the inconsistency of human knowledge, but it is entirely inconsistent to acknowledge the fundamental inconsistency of actual physical reality. Our physical reality, involving the universe/multiverse and quantum fields in it,  is produced by the need for a complete and consistent real \( \mathcal{F} \) in the Platonic realm. However, the application of Gödel's theorem on \( \mathcal{F} \) limits what can be obtained from  \( \mathcal{C} \), and it is not possible to obtain Gödelian consequences of \( \mathcal{F} \) through \( \mathcal{C} \). Assuming the existence of a non-computational part of reality \( \mathcal{N} \), in addition to the computational algorithm \( \mathcal{C} \), is the only way that reality can be consistent. Since \( \mathcal{N} \) can get past the Gödelian obstacles and even produce a consistent \( \mathcal{F} \) and \( \mathcal{C} \),   \( \mathcal{N} \) may actually be considered more fundamental than both \( \mathcal{F} \) and \( \mathcal{C} \), as it is capable of producing \( \mathcal{F} \) or \( \mathcal{C} \), but not vice versa. 

It may be noted that some ideas claim that the implications of \( \mathcal{N} \) have already been observed in nature. It has been suggested that the standard quantum mechanics based on Copenhagen interpretation has several problems, such as the need for an observer \cite{Sakurai, Ballentine, Nielsen}. These problems are resolved in a modification to quantum mechanics, where an objective collapse occurs \cite{Nature}. Furthermore, Copenhagen interpretation and most other interpretations of quantum mechanics need any exterior physical entity, so they can not be used to explain the quantum-to-classic transition in cosmology. This difficulty is again easily resolved by collapse models, where collapse occurs in an observer-independent and scale-dependent way \cite{j1, j2}.
An important approach to such objective collapse is based on gravitationally induced decoherence, where it is proposed that gravity plays a fundamental role in the decoherence of quantum systems, effectively acting as a mechanism that causes a quantum system to transition into a classical state. This has been studied using the Diosi-Penrose (DP) approach \cite{Physical, Physical1, Physical2, Physical3}, which postulates that quantum superposition of mass distributions gives a fundamental time-scale for decoherence. The approach holds that unstable superpositions between states with markedly different gravitational fields cause the system to collapse into one of the possible states \cite{Penrose1996}. This collapse time is proportional to the inverse of the gravitational self-energy of the difference between the mass distributions. Apart from this approach, other models of objective collapse have also been proposed \cite{ghirardi1986, Karolyhazy1966, Bassi2017}.
 It has been suggested that the fundamental indeterminacy in quantum mechanics in collapse models could be an example of a Gödelian phenomenon in physical theory \cite{Penrose1994}. The connection between quantum collapse and Gödel's theorem has been established in the Orch-OR theory \cite{or10, or1010}.  It has first been argued that consciousness in the human brain can overcome the Gödelian limitations due to the Lucas-Penrose argument. Then it is observed that the only physical theory, which is not computational is the measurement problem associated with quantum collapse. Thus, it has been argued in the Orch-OR theory, that consciousness in the human brain which produces non-algorithmic understanding is related to this quantum collapse. So, the 
Orch-OR theory uses such quantum collapse models to provide a basis for the original Lucas-Penrose argument (involving $F, C, N$). According to this theory, objective reduction of the brain’s quantum state creates consciousness, and consciousness is identified with the presence of $N$ in humans \cite{Penrose1994}. So, in the Orch-OR's description, quantum collapse which could be related to  \( \mathcal{N} \) provides a mechanism in the brain for the Lucas-Penrose contention that human cognition is non-algorithmic $N$ \cite{Penrose1989}. Thus, the mechanism underlies the Orch-OR  \cite{Penrose1994} is a Gödelian consequence obtained from \( \mathcal{F} \) via \( \mathcal{N} \) and not \( \mathcal{C} \), and this gives rise to $N$ in the human brain.
It may be noted that even if Orch-OR is not true, the argument in the  Platonic realm stands. This is because due to Gödel's theorems,  there will always exist 
consequences of \( \mathcal{F} \), which can only be obtained via \( \mathcal{N} \) and not \( \mathcal{C} \). It is possible that quantum collapse is such a consequence, but even if it is not, the existence of  \( \mathcal{C} \) is needed to overcome limitations imposed by the Gödel theorems.   So, a Gödelian consequence of  \(\mathcal{F} \) will be true, but it will only be possible to obtain it by \(\mathcal{N} \), and not \( \mathcal{C} \). Therefore, for any third quantized theory describing reality, we must have a non-algorithmic \( \mathcal{N} \) in the Platonic realm. This is the only way for it to be a fully consistent and complete description of reality. 

\section{Applications of this Third Quantized Theory}
Here, we discuss some interesting applications of this theory. We will demonstrate how the presence of \(\mathcal{N}\) is critically important for explaining important physical phenomena in the universe/multiverse. According to holography, the bulk AdS spacetime is dual to CFT on the boundary of that AdS spacetime \cite{ho, oh}. Now third quantization can explain the dynamical creation and annihilation of geometries, and hence topology-changing processes \cite{topol}. Thus, it is possible to construct processes that change asymptotically AdS spacetime to different geometry. Now it is possible that the information encoded in the boundary field theories could, in principle, capture all aspects of the bulk geometry, including such dynamical creation, annihilation, and topology-changing processes \cite{topol}. This suggests an interesting extension of the holographic principle, where the duality not only applies to single bulk geometries but also to superpositions and transitions between different geometric configurations. For instance, boundary states or entanglement structures could be interpreted as encoding the creation and annihilation of individual bulk spacetimes, while some nonlocal correlations could represent topological transitions.

Thus, the boundary dual to such a bulk superposition of spacetimes would be a combination of field theories, possibly with interactions or correlations dictated by the bulk dynamics, as 
\(
Z_{\text{boundary}} = \prod_i Z_{\text{CFT}_i} \prod_j Z_{\text{non-CFT}_j},
\)
where \(Z_{\text{CFT}_i}\) denotes the partition function of the \(i\)-th CFT dual to an AdS geometry and \(Z_{\text{non-CFT}_j}\) represents the partition function of the field theory corresponding to a non-AdS geometry.
The third quantization provides a natural way to study multi-geometric and multi-topological bulk configurations, and their intricate holographic duals would be a combination of field theories. The holographic reconstruction of bulk spacetimes from the boundary theories would need to account for such composite structures, suggesting the need for a more general understanding of holographic duality beyond the standard AdS/CFT correspondence. By encoding the dynamics of creation and annihilation of geometries, along with topology change in the third quantized bulk, and relating it to boundary quantum states, third quantization offers a novel avenue for exploring the full implications of the holographic principle in quantum gravity.

It is conceivable that black holes exist in the bulk, and even if we start from a spacetime without a black hole, there would be a finite probability of the creation of a black hole geometry in a third quantized theory of spacetime. It has been argued that holography can be used to address the information about the black holes in the bulk  \cite{ho12}.  Now it is possible that the information about these black holes can be encoded in the boundary theory in the form of Gödelian statements.  In this scenario, the information is preserved but cannot be algorithmically obtained within the formal system representing the boundary theory. If this perspective is correct, it offers a novel resolution to the black hole information paradox.
In conventional approaches to the black hole information paradox, two possibilities are typically considered: either the information is lost during black hole evaporation, violating unitarity \cite{hawking1976,banks1984, Polchinski:1994zs}, or it is preserved and retrievable within the framework of quantum gravity \cite{page1980,hawking2005,strominger1996}.
Here, a third possibility emerges: the information is preserved but fundamentally inaccessible within the boundary theory due to limitations imposed by Gödel’s incompleteness theorems  \cite{ho14}. Specifically, while the boundary theory may encode the information, it cannot be obtained through \(\mathcal{C}\), i.e., any logical derivation or algorithm within the boundary's formal system \(\mathcal{F}\). However, such information exists due to non-algorithmic understanding \(\mathcal{N}\) in the Platonic realm that transcends the Gödelian limitations of \(\mathcal{C}\).

Another possible application of this approach is to the paradoxes created by closed timelike curves. Such closed timelike curves exist in Gödel black holes  \cite{Kerner:2007jk,Srikanth:2007vz,Ren:2010zzb,Li:2012ee,Pourdarvish:2014hza,Pourhassan:2018scc}, which are solutions to Einstein's equations in general relativity.  Now in a third quantized theory, there is a finite (even if small) probability for any physically allowed geometry to form.  However, they allow for the possibility of time loops, where an event can influence its past. This creates self-referential paradoxes, such as the ``grandfather paradox" \cite{Morris:1988tu}. It is important to avoid such self-referential paradoxes in the universe/multiverse. Now we observe that in a formal system, Gödel's incompleteness theorems similarly arise from the self-reference within that system. Such self-referential problems that give rise to Gödel's incompleteness theorems can be addressed using   \({N}\). Here, again, we observe that using 
\({N}\), it can be directly observed that such solutions should not be considered physical as they would lead to self-referential paradoxes. This has led to the chronology protection conjecture, where it has been argued that the back reaction due to quantum effects prevents closed timelike curves from appearing in the universe \cite{H}.
The conjecture has not been explicitly proven, and the arguments are constructed to avoid self-referential paradoxes. However, the identification of such paradoxes and the need for their resolution have been based on
\(N\) rather than \(C\). 
Since this has been done by humans using \(N\), the absence of such paradoxes in the real universe/multiverse can occur due to the existence of  \(\mathcal{N}\) in the   Platonic realm.  Here, we again distinguish better the human understanding of the absence of such paradoxes, which occurs due to \(N\) in the human brain, and the actual absence of such paradoxes in the universe/multiverse, which occurs due to  \(\mathcal{N}\) in the Platonic realm. Thus,  the absence of such paradoxes could also possibly be explained by \(\mathcal{N}\) in the Platonic realm.
Furthermore, as we require full consistency, with zero probability (not a very small one) for such inconsistencies,  and there is finite (even if small) probability for any physically allowed geometry to form in a third quantized theory of gravity, it seems that for the full third quantized gravity, we have to resort to the Novikov self-consistency principle   \cite{Carlini:1995st,Carlini:1996ay}. 
It has been proposed, using the Novikov self-consistency principle in general relativity, that in spacetimes with closed timelike curves, the only solutions to the laws of physics that can occur are those that are self-consistent  \cite{Friedman:1990xc}. However, as this is a self-referential system, this self-consistency seems to have been argued using \({N}\) rather than \({C}\). This again implies that this may be actualized in the universe/multiverse by \(\mathcal{N}\) rather than \(\mathcal{C}\) in the Platonic realm.

However, we would like to point out that even if these specific examples could be explained by \(\mathcal{C}\), the general argument developed in this paper would stand, and we would still need \(\mathcal{N}\) in the Platonic realm for consistency and completeness. It may also be noted that even if quantum gravity uses a formalism different from third quantization, it would still be a formal structure in the Platonic realm from which spacetime would emerge due to a calculational algorithm \(\mathcal{C}\), again operating in the Platonic realm \cite{mi12, mi14}. Thus, the same argument would hold, and we would require a non-algorithmic understanding in the Platonic realm \(\mathcal{N}\) to overcome Gödelian limitations of \(\mathcal{C}\), and have a complete, consistent description of reality \cite{ho14}. Therefore, the main argument of this paper would hold even if quantum gravity is ultimately described by a formal structure different from a third quantized theory.

\section{Conclusion}
In this paper, we have argued that in any theory of quantum gravity, spacetime and quantum fields on spacetime would be an emergent structure, and should be dynamically produced. Thus, this theory should be constructed in the configuration space of fields rather than spacetime. Any such theory of quantum gravity would be a third quantized theory. In fact, the third quantized LQG can be represented by   GFT, and the third quantized string theory can be represented by SFT. It may be noted that SFT apart from being a second quantized theory of strings can also be viewed as a third quantized theory of conformal fields, and hence can be consistently analyzed as a third quantized theory. We use the axioms of string field theory to motivate the construction of such axioms for GFT. Then we use the general structure of axioms of both SFT and GFT, to construct the general feature of axioms for any third quantized theory of quantum gravity. As we have argued that such theories produce spacetime, they cannot be defined in spacetime. They rather exist in a Platonic realm, and spacetime emerges from them from a computational algorithm. Thus, apart from the formal axiomatic structure of the third quantized quantum gravity, a computational algorithm also exists in the Platonic realm. This actualizes the corollaries of that axiomatic system, and the universe/multiverse with quantum fields exits as a corollary of that axiomatic system. 

However, as it is a formal axiomatic system, Gödel theorem will apply to it. There will be things that are true but cannot be obtained from a computational algorithm. The consistency of the axiomatic system will be one such thing, which cannot be obtained from it. To overcome this difficulty, it is proposed that apart from the computational algorithm, it will also be possible to obtain non-computational non-algorithmic truths in the Platonic realm related to the axiomatic system. This is done by generalizing the original Lucas-Penrose argument to the Platonic realm. The main difference between the argument here and the original Lucas-Penrose argument is that the original Lucas-Penrose argument applies to human understanding of reality, and the argument here applies to the actual reality in the Platonic realm. This seems to be the only way to overcome the Gödelian limitations in the Platonic realm and produce a complete consistent third-quantized theory of quantum gravity.

\end{document}